\documentclass[a4paper,11pt]{article}
\usepackage{graphicx}

\def\spose#1{\hbox to 0pt{#1\hss}} %
\def\Dt{\spose{\raise 1.5ex\hbox{\hskip3pt$\mathchar"201$}}}    
\def\dt{\spose{\raise 1.0ex\hbox{\hskip2pt$\mathchar"201$}}}    

\def\deg{$^\circ$}

\def\lta{\mathrel{\spose{\lower 3pt\hbox{$\mathchar"218$}}
     \raise 2.0pt\hbox{$\mathchar"13C$}}}

\def\gta{\mathrel{\spose{\lower 3pt\hbox{$\mathchar"218$}}
     \raise 2.0pt\hbox{$\mathchar"13E$}}}

\begin{document}

\title{Exchange of Ejecta between Telesto and Calypso: \\
Tadpoles, Horseshoes, and Passing Orbits}
\maketitle\

\begin{center}

\author{Anthony R. Dobrovolskis \\
UCO/Lick Observatory, University of California, Santa Cruz, CA \\
and Space Science \& Astrobiology Division,
NASA Ames Research Center, MS 245-3, Moffett Field, CA 94035-1000}

\vspace{11pt}
\author{Jos\'e Luis Alvarellos \\
Space Systems/Loral, 3825 Fabian Way, MS G-76, Palo Alto, CA 94303}

\vspace{11pt}
\author{Kevin J. Zahnle and Jack J. Lissauer \\
Space Science \& Astrobiology Division,
NASA Ames Research Center, MS 245-3, Moffett Field, CA 94035-1000}

\end{center}

\begin{itemize}
\item 28 pages
\item 5 figures
\item 5 tables
\end{itemize}

\clearpage

\begin{verse}
\bf{Prepared for Icarus}\\
\bf{Submitted 5 Jan 2010}\\
\bf{Revised 2010 April 27}\\
\vspace{15pt}

\bf{Proposed Running Head:} Exchange of Ejecta between Telesto and Calypso 
\vspace{15pt}

\bf{Editorial correspondance, proofs and reprints:} \\

Anthony R. Dobrovolskis\\
245-3 NASA Ames Research Center\\
Moffett Field, CA 94035-1000\\
(650) 604-4194 \\
anthony.r.dobrovolskis@nasa.gov
\end{verse}

\clearpage

\begin{abstract}

We have numerically integrated
the orbits of ejecta from Telesto and Calypso,
the two small Trojan companions of Saturn's major satellite Tethys.
Ejecta were launched with speeds comparable to or exceeding their
parent's escape velocity, consistent with impacts into regolith surfaces.
We find that the fates of ejecta fall into several distinct categories,
depending on both the speed and direction of launch.

The slowest ejecta follow sub-orbital trajectories
and re-impact their source moon in less than one day.
Slightly faster debris barely escape their parent's Hill sphere
and are confined to tadpole orbits, librating about Tethys' triangular
Lagrange points $L_4$ (leading, near Telesto) or $L_5$ (trailing, near Calypso)
with nearly the same orbital semi-major axis as Tethys, Telesto, and Calypso.
These ejecta too eventually re-impact their source moon,
but with a median lifetime of a few dozen years.
Those which re-impact within the first ten years or so
have lifetimes near integer multiples of 348.6 days (half the tadpole period). 

Still faster debris with azimuthal velocity components $\gta$ 10 m/s
enter horseshoe orbits which enclose both $L_4$ and $L_5$ as well as $L_3$,
but which avoid Tethys and its Hill sphere.
These ejecta impact either Telesto or Calypso at comparable rates,
with median lifetimes of several thousand years.
However, they cannot reach Tethys itself;
only the fastest ejecta, with azimuthal velocites $\gta$ 40 m/s,
achieve ``passing orbits'' which are able to encounter Tethys.
Tethys accretes most of these ejecta within several years,
but some 1\% of them are scattered either inward to hit Enceladus
or outward to strike Dione, over timescales on the order of a few hundred years.

\end{abstract}


\clearpage

\section{Introduction}

There is a sense of youthfulness to Saturn's system,
as if the system were now being caught in the process
of evolving to something quieter. Signs of youth abound, 
each of which has something to tell us about what happened at Saturn:
(i) Saturn's rings appear both by color and dynamics to be tens or hundreds
of millions of years old, rather than the billions of years that describe the
Solar System. (ii) Enceladus' eruptions are too energetic to have been going on
as they are now for more than a small fraction of the age of the Solar System.
(iii) The methane in Titan's atmosphere will be chemically destroyed
on a time scale of order 10$^8$ yrs, with the hydrogen escaping to space;
there is no known source of new methane, nor is there any sign
on the surface of the hundreds of meters of photochemical product
which would have accumulated had methane been supplied to the atmosphere
over the age of the Solar System. 
(iv) Saturn's satellite system is crowded and has at least three examples 
of coorbital moons, a phenomenon not seen elsewhere in the Solar System.

One set of coorbital satellites consists of the two small moons 
Janus and Epimetheus ($\sim89$ km and $\sim59$ km in mean radius, respectively), 
traveling on mutual ``horseshoe'' orbits. The other two sets
each consist of a pair of small Trojan companions of larger moons:
Tethys is accompanied by Telesto (Smith et al. 1980)
and Calypso (Pascu et al. 1980; Seidelmann et al. 1981),
while Dione is accompanied by Helene (Lecacheux et al. 1980)
and Polydeuces (Porco et al. 2007).
Telesto, Calypso, and Helene are all 11-17 km in radius,
and were discovered by ground-based observations,
while tiny Polydeuces is only 1.3 km in radius
and was found with the Cassini orbiter (Porco et al. 2004).
From other spectacular Cassini images, 
Telesto and Calypso appear to be covered smoothly with fine material. 

All of these Trojan moons are intriguing 
because Tethys and Dione are adjacent; furthermore, 
Tethys is in a 4:2 mixed-inclination-type mean-motion resonance with Mimas, 
while Dione, Helene, and Polydeuces are all in a 2:1 inner-eccentricity-type
mean-motion resonance with Enceladus. The overall impression is paradoxical.
Small number statistics precludes surety,
but it seems unlikely that chance alone can explain
why Dione and Tethys each have two Trojans,
while all other major satellites have no such attendants.
Rather, there is a sense of underlying order,
as if all of this came about in a perfectly ordinary way.

The better to characterize this system, 
in this study we simulate the orbital evolution of test particles
launched from Tethys' coorbital satellites Telesto and Calypso. 
First we describe the methodology of the numerical integrations
and how we prepare the initial conditions.
Next we discuss evolution and fates of ejecta from Telesto and Calypso. 
The results turn out to be rather interesting: Both the speed and direction
of ejecta launch are important, with well-defined regimes
which correspond to tadpoles, horseshoes, and passing orbits.
The key points are best illustrated by strongly reductionist
numerical experiments in which ejecta are launched vertically into space
at various places along a satellite's equator.
We then broaden our compass to consider more realistic distributions
of ejecta launched from impact craters.
Our results imply that Telesto and Calypso readily exchange ejecta. 

\section{Initial Conditions and the Integration Model}

\subsection{Integrator}
We use the SWIFT integrator package (Levison and Duncan, 1994)
to evolve the Saturn system and the orbits of the ejecta forward in time.
SWIFT, which is based on work by Wisdom and Holman (1991),
is able to propagate the orbits of the massless test particles
as well as the massive bodies using a choice of integrators. For this study
we use its Regularized, Mixed Variable Symplectic integrator (RMVS3);
one of its advantages is the ability to use relatively large time-steps.
Another advantage of the RMVS3 method is that it can handle
very close approaches between a test particle and a massive body,
which are of great interest for this study.

We use a customized version of SWIFT, 
which we have modified to detect collisions with Saturn's rings 
and to include Saturn's optical flattening, 
along with its dynamical oblateness up to order $J_{12}$ 
(Dobrovolskis and Lissauer, 2004). 
All other massive bodies are treated as spherical; 
although the Trojan moons and Hyperion 
are some tens of percent out of round, so that their surface gravity 
and escape speed should vary by a comparable amount over their surfaces, 
this variation is minor compared with the speeds of their impactors and ejecta. 

\subsection{Massive Bodies Initial Conditions}
In our ejecta integrations we include
the gravitational attractions of Saturn, Mimas, Methone, Enceladus,
Tethys and its coorbitals Calypso and Telesto,
Dione and its coorbitals Helene and Polydeuces,
Rhea, Titan, Hyperion, Iapetus and Phoebe.
The initial state vectors and gravitational constants are from R. Jacobson
(sat263 model, pers. communication) in the form of barycentric initial locations
and velocities of the Saturn system for Epoch 26-JUL-2006, 00:00:00.0,
expressed in the ICRF system (Arias et al. 1995).\footnote{ICRF and J2000 
are almost identical coordinate systems; the ICRF celestial pole 
is offset from the mean pole at J2000 by less than 20 milliarcseconds.}
In addition, we modified SWIFT to include the gravitational perturbations of
the Sun by treating it as a special massive, distant ``satellite'' of Saturn;
we also added the mass of Jupiter to that of the Sun (Danby, 1962).
Initial conditions for the Sun and Jupiter were obtained from the DE200
JPL analytical Ephemeris. Finally, a rotation and a translation
were performed to align the fundamental plane with Saturn's equator
and to set the origin at the center of Saturn for our SWIFT integrations.

The initial state vector for Polydeuces is from Jacobson et al. (2008);
however, its J2000 barycentric initial conditions are for the Epoch 2-JAN-2004,
00:00:00.0. Hence, the previous initial conditions for the Saturn system
(minus Polydeuces) were integrated backwards in time using SWIFT
until 2-JAN-2004, at which point we ``drop'' the initial conditions
for Polydeuces (rotated and translated as well, of course)
into our final, ``augmented'' Saturn system to be used in the integrations.
Table 1 gives an overview of the Saturn satellite system used
in our integrations. Radii and density ($\rho$ = 0.6 g/cm$^3$)
of the coorbitals are from Porco et al. (2007).

\newpage
\begin{quotation}
\noindent
Table 1. Overview of the Saturn system used in this paper.
The values of $Gm$ and $R$ are from R. Jacobson (sat263 model),
except for Methone and the coorbitals, for which the values of $Gm$
were computed assuming a density of 0.6 g/cm$^3$ (Porco et al. 2007).
Orbital elements are mean values from Murray and Dermott (1999), 
except for Methone and Polydeuces, which are from Jacobson et al. (2008). 
Inclinations are given relative to the local Laplace plane. The semi-major axis 
values are shown in kilometers as well as in terms of Saturn's radius $R_S$. 
For Saturn, we use $GM = 3.7931206 \times 10^7$ km$^3$/s$^2$;
$R_S$ = 60,330 km; $J_2 = 16297 \times 10^{-6}$; $J_4 = -910 \times 10^{-6}$;
see Dobrovolskis and Lissauer (2004) for $J_6$, $J_8$, $J_{10}$ and $J_{12}$.

\end{quotation}
\begin{center}
\begin{tabular}{|l||c|c|c|c|c|c|c|}
\hline
          & $Gm$                & $R$ &   $a$   &  $a$  &  $e$  & $i$  & $P$  \\
Satellite & (km$^3$/s$^2$)      & (km)&   (km)  &($R_S$)&       &(\deg)&(days)\\
\hline
Mimas     & 2.5                 & 198 & 185,520 & 3.08  & .0202 & 1.53 & 0.942\\
Methone   & $6.7\times 10^{-7}$ & 1.6 & 194,230 & 3.22  & .0000 & 0.01 & 1.011\\
Enceladus & 7.2                 & 252 & 238,020 & 3.95  & .0045 & 0.02 & 1.370\\
Tethys    & 41.2                & 533 & 294,660 & 4.88  &   0   & 1.09 & 1.888\\
Calypso   & $2.0\times 10^{-4}$ & 10.6& 294,660 & 4.88  &   0   &  0   & 1.888\\
Telesto   & $3.2\times 10^{-4}$ & 12.4& 294,660 & 4.88  &   0   &  0   & 1.888\\
Dione     & 73.1                & 562 & 377,400 & 6.26  & .0022 & 0.02 & 2.737\\
Helene    & $7.3\times 10^{-4}$ & 16.5& 377,400 & 6.26  & .0022 & 0.02 & 2.737\\
Polydeuces& $3.7\times 10^{-7}$ & 1.3 & 377,200 & 6.25  & .0192 & 0.18 & 2.737\\
Rhea      & 153.9               & 764 & 527,040 & 8.74  & .0010 & 0.35 & 4.518\\
Titan     & 8978.1            &2776 & 1,221,850 & 20.25 &.0292 & 0.33 & 15.945\\
Hyperion  & 0.4               & 133 & 1,481,100 & 24.55 &.1042 & 0.43 & 21.277\\
Iapetus   & 120.5             & 736 & 3,561,300 & 59.03 &.0283 & 7.52 & 79.330\\
Phoebe    & 0.5               & 107 &12,952,000 & 215   & .163 & 175.3 &550.48\\
\hline
\end{tabular}
\end{center}

\subsection{The dynamical environment}
After an actual impact, the majority of ejected particles
stay relatively close to the moon's surface, following
suborbital trajectories and coming back rather quickly
to form the crater ejecta blanket and/or secondary craters.
However, in this study we address only those ejecta
whose speeds allow them to escape the source moon, i.e.,
those most energetic ejecta which are able to achieve saturnicentric orbits.
In the two-body problem, a particle escapes to infinity if its ejection speed
$V_0$ exceeds the classical escape velocity $V_e = \sqrt{2 Gm/R}$,
where $m$ and $R$ are respectively the mass and radius of the source body.
But because of three-body effects, a particle ejected from a moon
may be considered to escape if it reaches the satellite's Hill sphere.
The escape criterion then becomes $V_0 \geq V^*_e = \gamma V_e$
(Alvarellos {\it et al.} 2002, 2005), where the dimensionless correction factor

\begin{equation}
        \gamma = \sqrt{ \frac{1 - \chi}{1 - \chi^2 \sin^2 \zeta} } < 1
\end{equation}

\noindent
depends on both $\zeta$, the inertial angle of ejection
measured from the vertical, and $\chi \equiv R/R_H$, the ratio
of the satellite's physical radius $R$ to its Hill radius $R_H$, defined as

\begin{equation}
            R_H \equiv a [\mu/3]^{1/3}
\end{equation}

\noindent
where

\begin{equation}
            \mu \equiv \frac{m}{M+m}
\end{equation}

\noindent
and $m$ and $M$ are the masses of the satellite and Saturn respectively.
See Table 2 for the Hill radii
and other properties of Tethys and its Trojan companions.

\begin{quotation}
\noindent
Table 2. Additional properties for the ``Tethys system''.
Here we tabulate the satellites' Hill radii $R_H$,
the ratio $\chi \equiv R/R_H$, their classical escape velocities $V_e$,
the correction factor $\gamma$ for $\zeta$ = 0, $45^\circ$, and $90^\circ$
from Eq. (1), and the acceleration of gravity
$g = Gm/R^2 = V_e^2/(2R)$ at the satellites' surfaces.
Note that we have assumed perfect spheres in all three cases.
\end{quotation}
\begin{center}
\begin{tabular}{|l|c|c|c|c|c|c|c|}
\hline
         & $R_H$ & $\chi$ & $V_e$ &$\gamma_0$&$\gamma_{45}$&$\gamma_{90}$& $g$\\
Satellite& (km)  &        & (m/s) &          &             &       &(cm/s$^2$)\\
\hline
Tethys   & 2100  & 0.253  &  393  &   0.864  &    0.878    & 0.893 & 14.5  \\
Telesto  & 41.6  & 0.298  &  7.18 &   0.838  &    0.857    & 0.878 & 0.208 \\
Calypso  & 35.6  & 0.298  &  6.14 &   0.838  &    0.857    & 0.878 & 0.178 \\
\hline
\end{tabular}
\end{center}

As Table 2 shows, $\gamma$ depends weakly on $\zeta$.
Note that Kempf {\it et al.} (2010, section 3) give a formula
equivalent to $\gamma = \sqrt{1-\chi}$; this is tantamount to Eq. (1)
for the special case of vertical ejection ($\zeta$ = 0).
In any case, when $\chi$ is small,
$\gamma$ tends to unity and the classical result is recovered;
but if $\chi$ approaches unity, $\gamma$ becomes small.
The approximation that the ejectum escapes if it reaches the Hill sphere
is really a simplification of the full problem;
see Dobrovolskis and Burns (1980).

Ejecta particles approaching and leaving the source moon's Hill sphere
encounter a somewhat complicated dynamical environment.
A simple model which describes the dynamics rather well
is the circular restricted three-body problem,
where there are two bodies with finite masses (Saturn and Tethys)
which revolve about their common center of mass in circular orbits
due to their mutual gravitational attraction;
a test particle with infinitesimal mass is gravitationally influenced by
(but does not itself influence) the motion of the two massive bodies;
see Szebehely (1967) or Murray and Dermott (1999) for details.

Consider now a rotating frame such that the \textit{x}-axis passes
through the center of the two massive bodies: Lagrange
found that the equations of motion of the test particle in such a frame
exhibit five equilibrium points. Three of these, $L_1$, $L_2$ and $L_3$,
lie along the rotating \textit{x}-axis ($L_1$ and $L_2$ on either side
of Tethys, while $L_3$ lies almost opposite Tethys from Saturn),
and the $L_4$ and $L_5$ points form equilateral triangles
with Saturn and Tethys (see Fig. 1). The collinear points are always unstable,
but the equilateral points are stable for the Saturn-Tethys case ($\mu \ll 1$).
Note that the Trojan moons are not quite at rest at these stable points;
Telesto librates about $L_4$ with an amplitude of $\sim2^\circ$,
while Calypso librates about $L_5$ with an amplitude of $\sim4^\circ$.
The shape of their paths in the rotating frame
gives rise  to their description as ``tadpole'' orbits.
Larger librations which enclose both stable points
as well as the unstable $L_3$ point are called ``horseshoe'' orbits.

\begin{figure}
\begin{center}
\includegraphics[width=160mm]{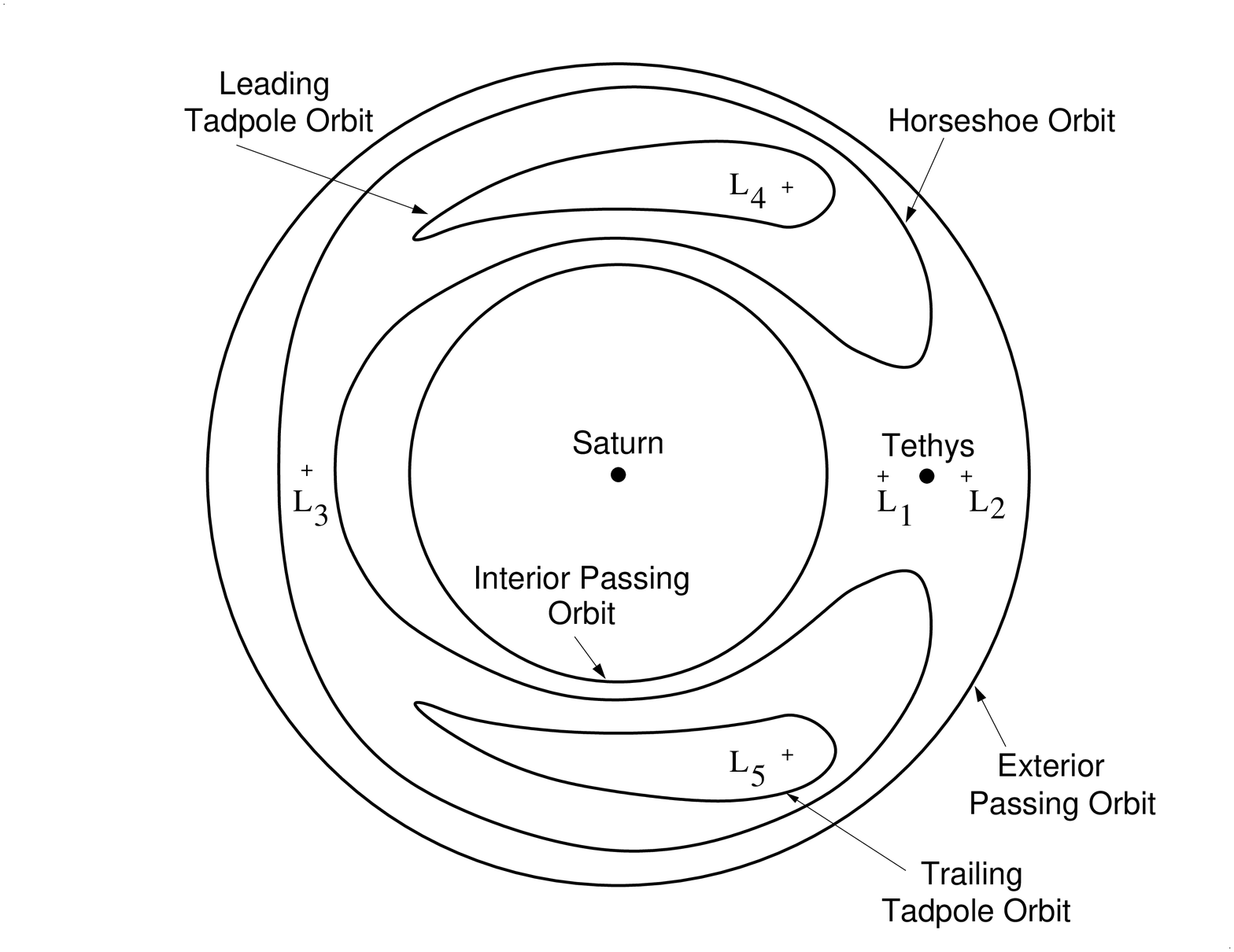}
\caption{Schematic of orbit classes in the rotating frame 
in which Saturn and Tethys are both fixed. 
Figure courtesy of Pat Hamill. See text. }
\end{center}
\end{figure}

\section{Fates of ejecta from the coorbitals}

In 2000-year-long simulations of ejecta escaping from Tethys,
Alvarellos et al. (2005; see also Dobrovolskis et al. 2007)
found that it takes between 650 and 1180 years to remove 99\%
of the test particles. Therefore at first we set the duration
of our simulations of ejecta from Tethys' co-orbitals to the same 2000-yr limit.
However, we soon found that as many as 30\% of the test particles
were still orbiting Saturn after this span of time.
A quick check then showed that
over 99\% of the debris was removed from the system
when we extended our integrations to 100,000 years.
Hence, effective removal of escaping ejecta from a coorbital
takes approximately two orders of magnitude longer than ejecta from Tethys.

To understand further the dynamical environment experienced by ejecta
from the coorbital satellites, we set up the following numerical experiment.
According to Eq. (1), the minimum speed at which we expect to see test particles
escaping from Calypso is $v^*_{esc} \approx$ 5.4 m/s. From the subsaturn point
on Calypso's equator (that is, longitude = 0, latitude = 0), we launch 
fifty particles radially outwards with speeds ranging from 5 to 35 m/s.
(We chose a slightly lower limit to allow for some margin;
the upper limit is set somewhat arbitrarily).
These initial conditions are rotated and translated
to the planetocentric coordinate system of the massive bodies.
We also launch another 50 particles in the same fashion,
but this time from a point at a longitude of $30^\circ$ along the equator,
and the same number every 30 degrees along Calypso's equator
until we reach 330\deg, for a total of 600 particles.
Then we integrate the system forward in time, and record the results.

\newpage
\clearpage
\begin{figure}
\begin{center}
\includegraphics[scale=0.6]{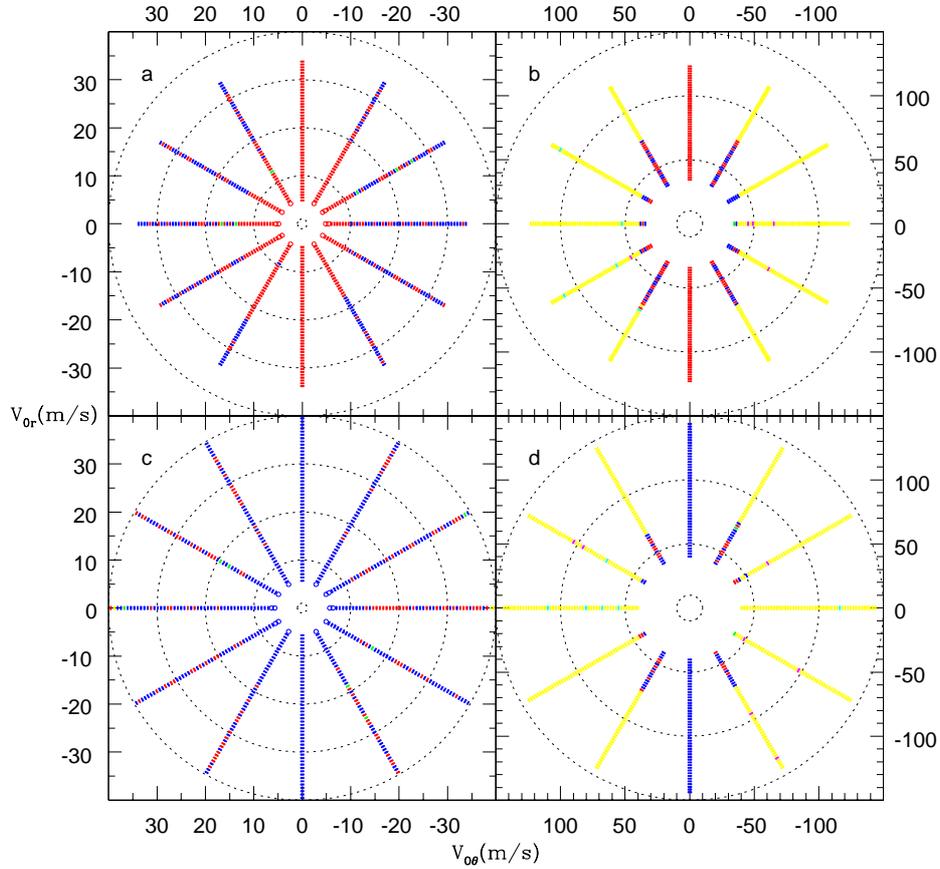}

\caption{ The fates of circumequatorial ejecta from Telesto and Calypso. 
Panel a (top left): slow ejecta from Calypso (5 m/s $\le V_0 <$ 34 m/s). 
Panel b (top right): fast ejecta from Calypso (34 m/s $< V_0 \le$ 123 m/s). 
Panel c (bottom left): slow ejecta from Telesto (6 m/s $\le V_0 <$ 40 m/s). 
Panel d (bottom right): fast ejecta from Telesto (40 m/s $< V_0 \le$ 144 m/s).
The vertical scale in each panel represents the radial component $V_{0r}$ 
of the launch speed from the satellite's surface, 
so that Saturn is toward the bottom of the figure, 
while the horizontal scale represents its azimuthal component $V_{0\theta}$; 
note that $V_{0\theta}$ increases to the left, 
in the direction of the apex of orbital motion. In polar coordinates, 
the angle variable represents the longitude of launch along Calypso's equator, 
and the radial coordinate represents the ejection speed $V_0$. 
The dotted circles in each panel represent launch speeds of 1, 10, 20, 30, 
and 40 m/s (panels a and c), or of 10, 50, 100, and 150 m/s (panels b and d). 
RED symbols represent particles which hit Calypso, and 
BLUE ones denote particles which hit Telesto. 
YELLOW dots represent particles which hit Tethys, while
MAGENTA ones hit Enceladus, and 
CYAN ones hit Dione.
GREEN represents those particles still orbiting Saturn 
at the end of the simulation ($t$ = 100,000 years). 
The open symbols near the centers of panels a and c 
represent those particles which went into suborbital trajectories and 
came back to their source moon after only a few hours; all others are solid. }
\end{center}
\end{figure}
\newpage

Panel (a) of Fig. 2 
shows the fates of this first set of 600 circumequatorial ejecta. 
Several observations can be made:
The first one is that most of the slowest particles
never left Calypso's Hill sphere, 
but instead went into suborbital trajectories (open symbols). 
The second observation is that all particles
ejected towards the sub- and anti-Saturn directions
(that is, along the axis corresponding to $ V_{0r}$,
the component of the relative velocity away from Saturn)
come back to Calypso after spending some time orbiting Saturn.
The third observation is that most ejecta which did not come back to Calypso
went to Telesto rather than Tethys; in fact, no ejecta from this set hit Tethys!

Panel (b) of Fig. 2 shows the fates of 600 more particles, ejected from 
the same points as before along Calypso's equator, but at higher speeds. 
Again, those particles with zero azimuthal velocity component $V_{0\theta}$
all come back to Calypso. But many of these fast particles
with a non-zero $V_{0\theta}$ are able to travel to nearby Tethys, and a few 
in fact are seen even to reach distant places such as Enceladus and Dione.

Panels (c) and (d) of Fig. 2 show the equivalent plots for the other coorbital,
Telesto. The patterns are rather similar, as expected:
Particles ejected into the subsaturn and antisaturn directions
($ V_{0\theta}=0$) always came back to Telesto.
Otherwise, most slow ejecta hit either Telesto or Calypso,
while many fast ejecta reached Tethys, or even Enceladus or Dione.

In addition, we launched 100 particles (not plotted)
straight up from the north and south poles of both Telesto and Calypso.
Every one of these 400 polar ejecta eventually returned to its parent satellite,
just like particles ejected towards the sub- and anti-Saturn directions, 
except that three ejecta from the poles of Telesto survived the integration. 

Table 3 lists the actual numbers and percentages for the fates
of all the circumequatorial ejecta.
Calypso regained 45\% of its own ejecta;
similarly Telesto regained 49\% of its own.
These particles that came back to their source moon could have been on tadpole
or horseshoe orbits. Perhaps somewhat naively one would expect
the big moon Tethys to accumulate a very large fraction of coorbital ejecta;
in fact it obtained 31\% and 35\% of Calypso and Telesto
circumequatorial ejecta respectively. The caveat, however,
is that these numbers depend critically on the assumed velocity distribution.
Certainly as the ejection speeds get higher,
a larger fraction of ejecta ends up on Tethys (see Figs. 2b and 2d).

Is the transfer of ejecta from Calypso to Telesto equivalent
to the transfer from Telesto to Calypso? Surprisingly,
Telesto obtained 23\% of the Calypso ejecta, 
but Calypso got only 14\% of the Telesto ejecta: 
a ratio of 267/163 $\approx$ 1.64. 
In fact Telesto is slightly larger than Calypso, 
but the ratio of their geometric cross-sections $\pi R^2$ is only $\sim1.37$; 
hence it seems Telesto got more than its fair share.
With gravitational focussing, however, their collisional cross-sections 
scale between $R^2$ and $R^4$, for a ratio up to $\sim1.87$, consistent 
with our results. We will address these questions further in future work. 

\newpage
\begin{quotation}
\noindent
Table 3. Fate of circumequatorial ejecta from the coorbitals
after 100,000 years; the fast and slow sets have been combined.
$N$ is the number of particles. Percentages have been normalized to the number
of particles which entered Saturn orbit (labeled ``Total'').
``Active'' means that the particles were still orbiting Saturn
at the end of the integration; 
all of these survivors were still in tadpole or horseshoe orbits.
\end{quotation}
\begin{center}
\begin{tabular}{|l||c|c||c|c|}
\hline
&\multicolumn{2}{c||}{\textbf{Ejecta from Calypso}}&\multicolumn{2}{c|}{\textbf{Ejecta from Telesto}}\\
\hline
\textbf{Target} & $N$   & Percent   & $N$  & Percent \\
\hline
Enceladus       & 5     & 0.4\%     & 7    & 0.6\% \\
Tethys          & 371   & 31.3\%    & 417  & 35.2\% \\
Calypso         & 531   & 44.8\%    & 163  & 13.7\% \\
Telesto         & 267   & 22.5\%    & 584  & 49.2\% \\
Dione           & 5     & 0.4\%     & 6    & 0.5\% \\
Active          & 7     & 0.6\%     & 9    & 0.8\% \\
\hline
Total           & 1186  & 100\%     & 1186 & 100\% \\
\hline
\end{tabular}
\end{center}

\section{Theory of ejecta from coorbitals}

In order to understand the above results in a more quantitative sense,
note that ejecta which escape from the Hill sphere
of Telesto or Calypso with a small relative speed
are also in tadpole orbits with respect to Tethys.
For graphic examples, see Figs. 3.15 and 3.16
on pages 96 and 97 of Murray and Dermott (1999).
As those figures show, such orbits may have significant eccentricities
and inclinations, but these are unimportant for our purposes.
In contrast, the semi-major axis and mean motion are crucial.

Now Telesto and Calypso are in nearly circular orbits
with essentially the same semi-major axis $a$ and mean motion $n$ as Tethys.
Then a particle ejected from either one with a relative velocity
$\Delta \bf{V}$ (after escape) suffers a change in semi-major axis of
\begin{equation}
                \Delta a \approx 2 \Delta V_{\theta} / n ,
\end{equation}
and a corresponding change in mean motion of
\begin{equation}
                \Delta n \approx -3 \Delta V_{\theta} / a
\end{equation}
(Danby 1962, p. 242), where $\Delta V_{\theta}$ is the {\it azimuthal} component
of $\Delta \bf{V}$, in the direction of increasing orbital longitude $\theta$.
The relative speed $\Delta V$ is found in the usual way (Bate et al., 1971), 
but with the classical escape speed replaced by the modified escape speed 
$V^*_e = \gamma V_e$: 

\begin{equation}
            \Delta V^2 = V_0^2 - (V^*_e)^2 ,
\end{equation}

\noindent
where $V_0$ is the ejection speed at the surface.
For small amplitudes, the angular frequency of the longitude libration
is $n \sqrt{27 \mu /4} \approx 1.04 \times 10^{-7} {\rm s}^{-1}$ (Murray
and Dermott 1999, p. 94), corresponding to a tadpole period of $\sim$697 days.
Now the amplitude of the particle's libration in $\theta$
grows with $|\Delta a|$ and $|\Delta n|$,
as demonstrated by Figs. 3.15 and 3.16 of Murray and Dermott (1999).
As the maximum and minimum azimuthal separation from Tethys
reach $180^\circ$ and $\sim24^\circ$, respectively, the tip of the 
tadpole orbit reaches the $L_3$ point (Sinclair 1984, Lissauer 1985).
Beyond this amplitude, the leading tadpole orbits enclosing $L_4$
and the trailing ones enclosing $L_5$ merge together into the family
of horseshoe orbits, which enclose $L_4$, $L_5$, and $L_3$ all at once,
but still avoid Tethys and its entire Hill sphere,
as exemplified by Fig. 3.17 on page 98 of Murray and Dermott (1999).

This tadpole/horseshoe transition occurs when
\begin{equation}
    \Delta a = \pm a \sqrt{8 \mu /3} \approx \pm 501 \, {\rm km}
\end{equation}
(Dermott and Murray 1981b; see also Murray and Dermott 1999, p. 106).
This critical amplitude defines the width of the tadpole zone,
and corresponds to a change in mean motion of
\begin{equation}
        \Delta n = - \frac{3}{2} \frac{n}{a} \Delta a
    = \mp n \sqrt{6 \mu} \approx \mp 9.83 \times 10^{-8} {\rm s}^{-1}.
\end{equation}
Note how this $\Delta n$ is only slightly less than
the tadpole frequency $n \sqrt{6.75 \; \mu}$ for small amplitudes.

Equating formulae (4) and (7), or (5) and (8),
and solving for $\Delta V_{\theta}$ then gives
\begin{equation}
\Delta V^{'}_{\theta} = \pm na \sqrt{2 \mu /3} \approx \pm 9.66 \, {\rm m/s},
\end{equation}
where the prime designates the critical value.

Now the above result gives the critical magnitude of the azimuthal velocity
{\it after} escape from the coorbitals of Tethys.
The corresponding azimuthal velocity at launch $V^{'}_{0 \theta}$
must be corrected for the escape speed from the parent moon:
\[
   \Delta V^{'}_{\theta} = V^{'}_{0 \theta} \sqrt{1 -(V_e^* / V_0)^2}
\]
\[
            \Longleftrightarrow
\]
\begin{equation}
 V^{'}_{0 \theta} = \Delta V^{'}_{\theta} \sqrt{1 +(V_e^* / \Delta V)^2} .
\end{equation}
Assuming that the ejecta are launched straight ahead or straight back
along the satellite's orbit, so that $V_0 = |V^{'}_{0 \theta }|$,
these formulae give $V^{'}_{0 \theta} \approx$ 11.38 m/s for Telesto, 
and 10.94 m/s for Calypso.

At slower azimuthal velocities than the above, ejecta from the coorbitals
are confined to tadpole orbits, and can only reimpact their parent moons.
Above these limits, the ejecta follow horseshoe orbits,
and may collide with either coorbital with roughly equal probability.
Horseshoe ejecta cannot collide with Tethys,
however, or even enter its Hill sphere.

Horseshoe orbits are confined in width, though, much like the tadpole orbits;
the width of the horseshoe zone is comparable to Tethys' Hill radius
$R_H \approx$ 2100 km (Dermott and Murray 1981a).\footnote{Izidoro 
{\it et al.} (2010) cite Dermott and Murray (1981a) as giving 
$a \mu^{1/3} /2 \approx 0.6934 R_H$ for the width of the horseshoe zone, 
but we cannot find support for this coefficient.} 
Setting $\Delta a = \pm R_H$ in formulae (5) and (4) yields a change 
in mean motion of $\Delta n \approx \mp 4.12 \times 10^{-7} {\rm s}^{-1}$, 
and a difference in azimuthal velocity of
\begin{equation}
\Delta V^{''}_{\theta} = n R_H / 2 \approx \pm 40.4 \, {\rm m/s}
\end{equation}
after escape from the parent moon, where the double prime designates
a second critical value. Correcting for escape speed then gives critical
launch speeds of $\sim$40.9 m/s from Calypso, and $\sim$41.1 m/s from Telesto.

For azimuthal velocities greater than the above magnitude,
ejecta from the coorbitals are no longer confined to horseshoe orbits, but 
instead follow ``passing'' orbits which are unrestricted in orbital longitude. 
Particles with $\Delta V^{''}_{\theta} \lta$ -41 m/s
($\Delta a < 0$, $\Delta n > 0$) can overtake Tethys from the ``inside track'';
similarly, those with $\Delta V^{''}_{\theta} \gta$ +41 m/s
($\Delta a > 0$, $\Delta n < 0$) can ``undertake'' Tethys from exterior orbits; 
in both cases, the relative synodic period is $< 2\pi/\Delta n \approx$ 176 d. 
In either case, such ejecta can penetrate Tethys' Hill sphere,
and eventually can hit Tethys itself, or can be scattered by a close encounter
into orbits which cross another satellite, such as Enceladus or Dione.

The above interpretations and the corresponding critical speeds
$V^{'}_{\theta} \approx$ 11 m/s and $V^{''}_{\theta} \approx$
41 m/s agree fairly well with the patterns seen in Fig. 2.
Note that the first critical value $\Delta V^{'}_{\theta}$
scales as the 1/2 power of $\mu$,
while the second critical value $\Delta V^{''}_{\theta}$ goes as its 1/3 power.
Both of these values depend on the mass of Tethys (and Saturn)
but are independent of the much smaller masses of the coorbital moons.
Furthermore, the horseshoe zone is wider than the tadpole zone
(and $\Delta V^{''}_{\theta} > \Delta V^{'}_{\theta}$)
provided that $\mu \lta 3/512 \approx$ 0.006 
(as seen by equating formulae 2 and 7). 

\section{Longitude libration}

\begin{figure}
\begin{center}
\includegraphics[width=160mm]{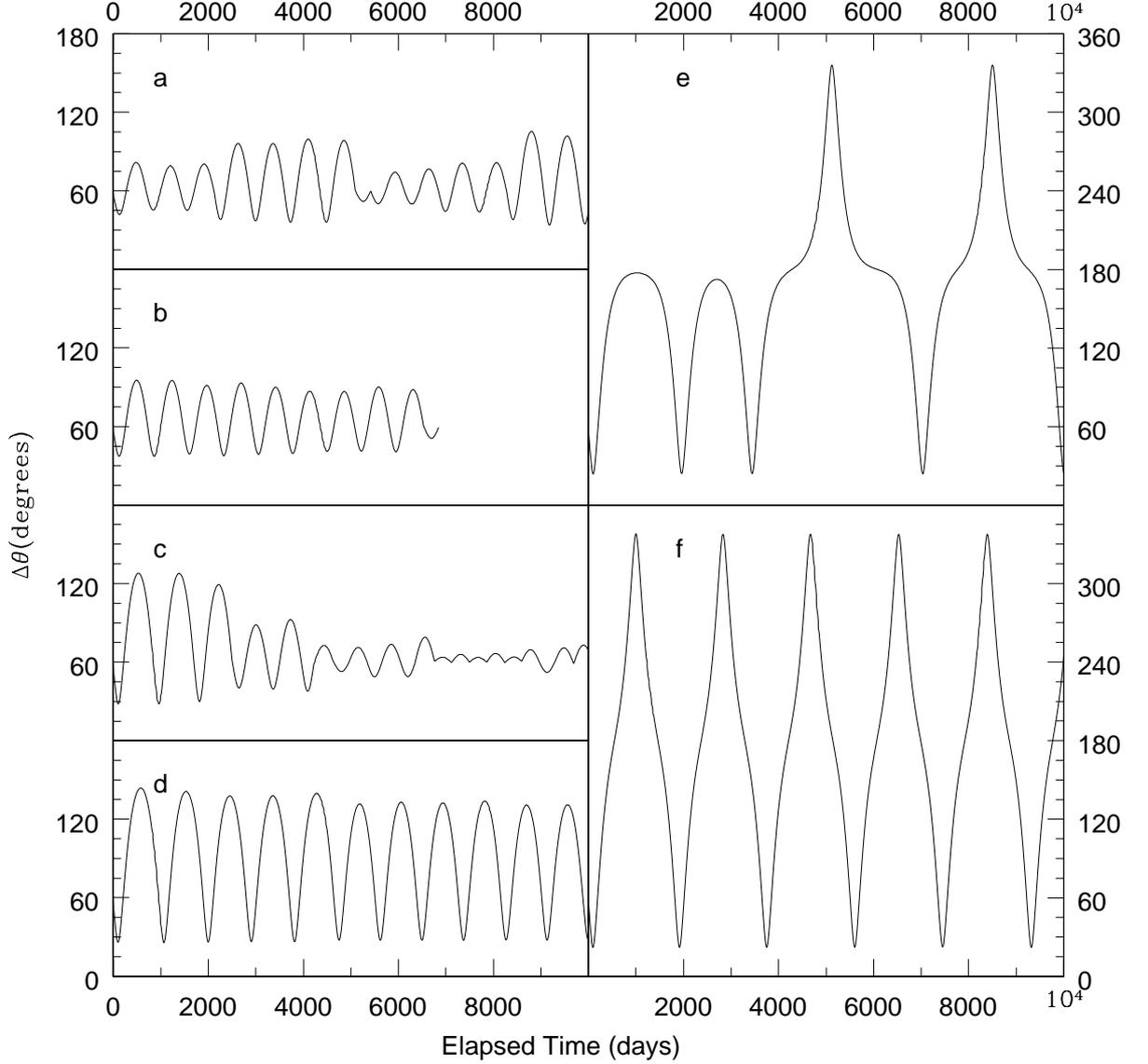}
\caption{Longitude librations of ejecta particles
launched with speed $V_0$ straight ahead from the apex of Telesto.
$\Delta \theta$ is the difference between the mean longitudes of Tethys
and of the particle.
Panel (a): particle 4, $V_0$ = 7.74 m/s.
Panel (b): particle 5, $V_0$ = 8.43 m/s.
Panel (c): particle 8, $V_0$ = 10.5 m/s.
Panel (d): particle 9, $V_0$ = 11.2 m/s.
Panel (e): particle 10, $V_0$ = 11.9 m/s.
Panel (f): particle 11, $V_0$ = 12.6 m/s. }
\end{center}
\end{figure}

In order to verify our interpretations,
we tracked the librations in longitude.
At each output timestep,
we calculated the difference $\Delta \theta \equiv \theta_p -\theta_T$
between the mean longitudes of Tethys and of each ejecta particle.
(Using the mean longitude instead of the true longitude
averages out the epicycles.)
For tadpole orbits near $L_4$ and Telesto,
$\Delta \theta$ librates around 60$^\circ$;
for those near $L_5$ and Calypso,
$\Delta \theta$ librates around 300$^\circ$.
Horseshoe orbits extend from $\lta 24^\circ$ to $\gta 156^\circ$
(Sinclair 1984, Lissauer 1985; see also Murray and Dermott 1999, p. 128).

Figure 3 plots the resulting $\Delta \theta$ for representative ejecta particles
launched at various (slow) speeds $V_0$ straight ahead from Telesto's apex of motion.
All six of the particles displayed eventually returned to their source
by colliding with Telesto again. Particle 4 (panel a, $V_0$ = 7.74 m/s)
repeatedly changes its libration amplitude, with a particularly close 
encounter with Telesto at $\sim$5170 days. In comparison, particle 5 
(panel b, $V_0$ = 8.43 m/s) performs a relatively uniform libration 
with an amplitude of $\sim30^\circ$ and a period of $\sim$700 days,
but suffers a close encounter with Telesto after 9 full periods
(note the slight kink in the curve), and then finally hits it
after another half-cycle. Likewise, particle 8 (panel c, $V_0$ = 10.5 m/s)
begins with a large libration amplitude of $\sim50^\circ$,
but its amplitude is repeatedly reduced until it suffers a whole series
of close encounters with Telesto after $\sim$7000 days.

In contrast, particle 9 (panel d, $V_0$ = 11.2 m/s)
executes a relatively uniform libration
with a very large amplitude of $\sim55^\circ$ and a period of $\sim$900 days.
Note how the tadpole period lengthens with amplitude,
and how this graph is noticeably non-sinusoidal,
because tadpole librations are not strictly harmonic. Particle 10
(panel e, $V_0$ = 11.9 m/s) performs two large, slow tadople librations,
but then at $\sim$4300 days it undergoes a transition to a horseshoe orbit
of long period. Finally, particle 11 (panel f, $V_0$ = 12.6 m/s)
embarks directly onto a horseshoe orbit of shorter period $\sim$1900 days.
Note how the transition velocity agrees with our prediction
of $\Delta V^{'}_{\theta 0} \approx$ 11.4 m/s, above.
These examples validate our interpretations, and demonstrate
the dependence of the tadpole/horseshoe transition on launch velocity.

\section{Ejecta lifetimes}

As further confirmation of our interpretations, we also examined 
the distribution of removal times of ejecta from our simulations. 
The overall lifetime distributions show significant structure, but are 
considerably clearer when partitioned according to the fates of the ejecta, 
as displayed in Fig. 4. 

Panels 4a and 4b respectively show histograms 
of the lifetimes of ejecta from Telesto and Calypso 
which were removed by collision with Tethys. 
Note that these lifetimes are on the order of only $10^1$ years. 
The corresponding particles were all on passing orbits, 
and encountered Tethys early and often before colliding with it. 
In contrast, those few ejecta in passing orbits scattered by Tethys into 
colliding with Enceladus or Dione (not shown) typically lasted a few centuries. 

For comparison, panels 4c and 4d respectively display the lifetimes 
of particles from Telesto removed by colliding with Calypso, 
and those from Calypso removed by Telesto. In contrast to panels 4a and 4b, 
these lifetimes are on the order of $10^4$ years! 
These particles were all on horseshoe orbits, 
and went through many librations before colliding with the other coorbital. 

Panels 4e and 4f respectively display the removal times of ejecta 
from Telesto and Calypso removed by collision with the source moon. 
These histograms show a combination of the long lifetimes associated 
with horseshoe orbits, and other lifetimes on the order of $10^1$ years. 
These latter do not correspond to passing orbits, however, 
as in panels a and b, but rather to tadpole orbits. 
Note the steep cutoff of this distribution below $\sim10^0$ years, 
in contrast to the left-hand tails at short times in panels a and b. 
Panels 4e and 4f also show a small detached peak at lifetimes 
on the order of $10^{-3}$ years, corresponding to suborbital ejecta. 

Finally, panels g and h of Fig. 4 show histograms of the removal times of ejecta
launched straight up from the poles of Telesto and Calypso, respectively. 
These ejecta were all reaccreted by their parent moon, with the exception 
of three particles from Telesto which survived the integration. 
All of the rest show a broad distribution of lifetimes centered about 
$\sim10^2$ years, unlike any of the other categories discussed above. 

\clearpage
\begin{figure}
\begin{center}
\includegraphics[width=130mm]{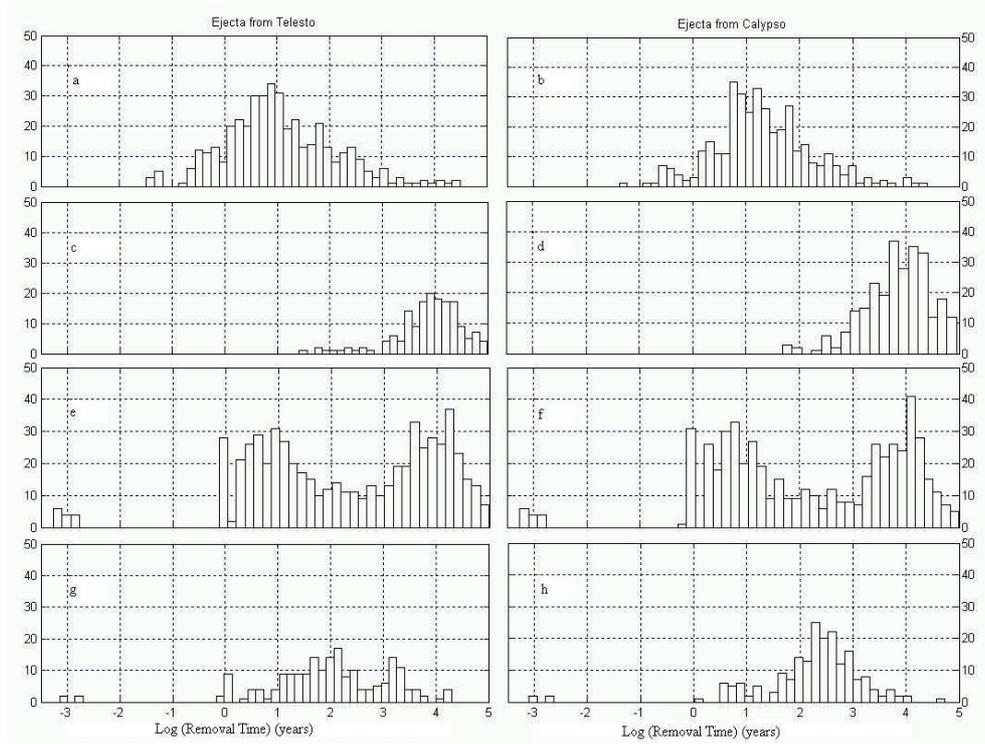}
\caption{Distribution of removal times for ejecta launched vertically 
from Telesto and Calypso, partitioned according to source and fate. See text. 
Panel a: Equatorial ejecta from Telesto, removed by collision with Tethys. 
Panel b: Equatorial ejecta from Calypso, removed by collision with Tethys. 
Panel c: Equatorial ejecta from Telesto, removed by collision with Calypso. 
Panel d: Equatorial ejecta from Calypso, removed by collision with Telesto. 
Panel e: Equatorial ejecta from Telesto, removed by collision with Telesto. 
Panel f: Equatorial ejecta from Calypso, removed by collision with Calypso. 
Panel g: Polar ejecta from Telesto, removed by collision with Telesto.
Panel h: Polar ejecta from Calypso, removed by collision with Calypso. }
\end{center}
\end{figure}

\clearpage

\subsection{Discretization of removal times}

\begin{figure}
\begin{center}
\includegraphics[width=130mm]{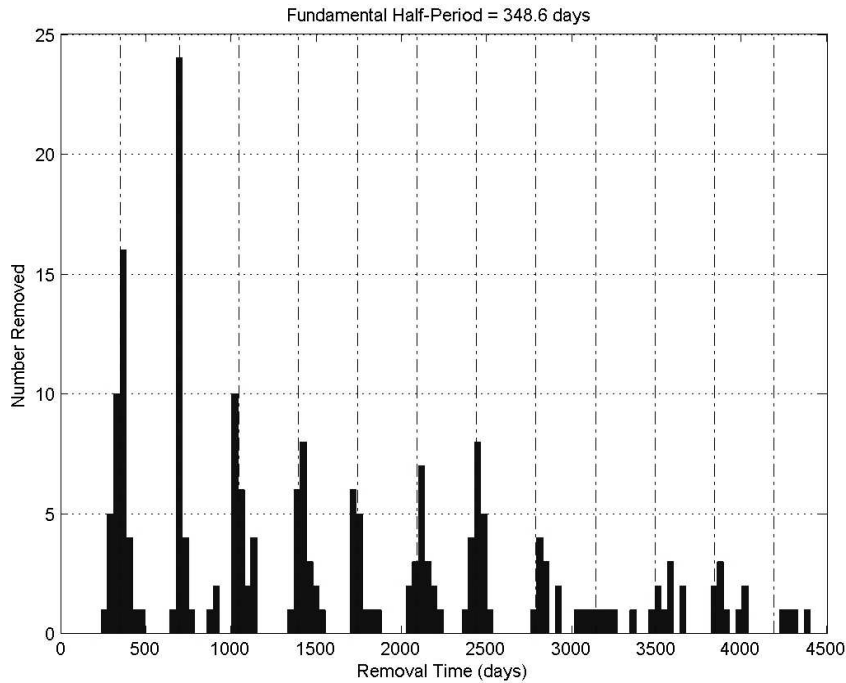}
\caption{Quantization of removal times for circumequatorial ejecta
(both Calypso and Telesto); only the relatively early removers show this effect.
Vertical lines denote integer multiples of 348.6 days, half the tadpole period. }
\end{center}
\end{figure}

Consider now the distributions of removal times shown in Fig. 4.
If we combine  particles removed from Calypso and Telesto
and zoom in on the escaping particles removed ``promptly'' (defined here 
as particles which took more than 1 day but less than 10 years to be removed)
in the removal times distribution plot, a curious phenomenon reveals itself.
As can be seen from Fig. 5,
the lifetimes of ejecta from these Trojan satellites are discretized,
with removals occurring in bursts at intervals of just less than a year.
An analogous plot for the polar ejecta (not shown) looks similar. 

The key to understanding such behavior is to recall that in a rotating frame,
the motion of an object in a low-amplitude tadpole orbit can be described
as a combination of two motions: a short-period circulation about an epicenter,
with period approximately equal to its orbital period, plus a longer-period
libration of the epicenter about the equilibrium point ($L_4$ or $L_5$)
with a period
\begin{equation}
        P_{\rm{tad}} \approx \frac{2 P_{\rm{orbit}}}{\sqrt{27 \mu}},
\end{equation}
provided that $\mu \ll 1$ (Murray and Dermott, 1999).
Here $P_{\rm{orbit}} \approx 1.888$ days is the orbital period of Tethys.

If we plug in the numbers from Table 1, 
we obtain a period of 697.2 days, or a half-period of 348.6 days;
integer multiples of this half-period are marked in Figure 5 as vertical lines.
It can be seen that the times of ejecta removal 
are clearly discretized near these values. As time goes on, however,
this discretization becomes fuzzier until it is no longer detectable
after approximately five full periods, as is apparent from Fig. 5.
The lower-speed ejecta from Telesto and Calypso which embark on tadpole orbits
are re-encountering their source moon after half of a libration period,
and again and again at roughly equal intervals
(with a dispersion on the order of ten percent),
until they are eventually reaccreted.
The libration histories from Fig. 2 also support this interpretation.

\section{Ejecta from coorbitals produced by centaur impacts}

To investigate a somewhat more realistic transfer of material from Calypso to
Telesto and/or Tethys and vice-versa, we simulated the ejection of particles due
to centaur impacts at selected locations, such that the resulting crater radii
are 1 km. (An impact resulting in a larger crater than this probably would
disrupt the coorbital catastrophically, a subject beyond our present scope.)
To compute the initial conditions for the ejecta, we take the center of a crater
as the origin of a topocentric coordinate system (Bate {\it et al.}, 1971),
on which we then construct twenty concentric ``ejection annuli''.\footnote{To 
simplify matters, we did not map the flat topocentric
coordinate system to the spherical surface of the satellite (\textit{i.e.},
the topocentric coordinate system is a flat plane tangent
to the spherical satellite and touching it at the center of the crater).}
The inner radius of the innermost ejection annulus
is approximately given by the impactor radius, while the outer radius
of the outermost annulus is set to the distance from the impact site
where the ejection speed drops below the escape speed from the satellite.
We then compute the radii $x_i$ of the twenty ejection annuli in such a way
that all have the same area. The actual ejection ring radii correspond
to the mass median of each annulus. Then, given each ejection radius
we compute the ejecta velocities according to the ``rubble'' model
(Housen et al., 1983; Alvarellos et al., 2005).

In the outer Solar System and in the Saturn system in particular,
the most important impactors are the centaurs, 
formerly known as ecliptic comets (Shoemaker and Wolfe, 1982; 
Smith et al., 1982, 1986; Zahnle et al., 1998). For these objects,
the impact speed $U$ as obtained using Zahnle \textit{et al.}'s (2001)
Monte Carlo algorithm is given by
\begin{equation}
    U \approx \sqrt{3} V_{orb} \left(1 + 0.9 \cos\beta \right)^{0.35}
\end{equation}
(Alvarellos et al. 2005), where $\beta$ is the impact site's angular distance
to the apex of motion and $V_{orb} = na$ is the satellite's orbital speed;
all else being equal, an impact at the apex point of a satellite
occurs at a higher speed (and lower zenith angle) and hence produces
a larger crater, which brings us to the topic of crater size.

After the impact process itself is complete, a bowl-shaped depression
called a transient crater is left at the target surface.
Given an impactor of diameter $d$ km and density $\rho_i$
hitting a target surface of density $\rho_t$ at a speed $U$ 
and at incidence angle $\zeta$ from the vertical,
the diameter of a simple/transient crater is given by
\begin{equation}
D_t = 1.1 d \left(\frac{U^2}{gd}\right)^{0.217} \left(\frac{\rho_i \cos\zeta}
{\rho_t}\right)^{1/3}
\end{equation}
(Zahnle \textit{et al.} 2003), where $g$ is the local acceleration of gravity. 
For small craters ($D_t \le$ 15 km),
the final crater diameter is equal to the transient crater diameter $D_t$.

The ejection of debris takes place immediately after the impact. 
Loose material expected to be found at the surfaces of the low-gravity Trojans 
is assumed to be launched from the surface at an angle of $45^\circ$ 
from the vertical, at speeds 
\begin{equation}
	        V_0 = K \sqrt{g R_t} (x/R_t)^{-e_x}
\end{equation}
(Housen et al., 1983); here $x$ is the distance from the impact point,
$R_t = D_t/2$ is the transient crater radius, and both $K \approx 0.57$ and
$e_x \approx 1.770$ are dimensionless constants (see Alvarellos et al. 2005).
The resulting ejection speeds range from 5 m/s to 30 m/s; 
launch speeds from the bigger Trojan Telesto are slightly higher than 
from Calypso, since $V_0$ scales as the square root of $g$. As before, 
these ejecta initial conditions are rotated and translated from the topocentric
to the planetocentric coordinate system for ingestion into the integrator.

We assumed an impact producing a 1-km-radius crater at five ``cardinal'' points 
of each Trojan moon (apex, subsaturn, antisaturn, north pole, and south pole). 
However, very little ejecta would actually escape from the antapex of motion 
in the rubble model, because of the low impact speeds there. 
For ejecta from the antapex, therefore, 
we used instead the ``spallation'' model of Melosh (1984; 1985a, 1985b; 1989), 
as modified by Zanhle et al (2008). In this case, 
the ejection speeds spanned the range 5 m/s $\lta V_0 \lta$ 56 m/s, 
so the fastest particles are injected into passing orbits and can reach Tethys. 
In fact, something like 5\% of these antapex spalls do reach Tethys, with most 
of the rest (between 80\% and 83\%) coming back to the source Trojan moon.

Table 4 lists the fates of ejecta particles ejected from 1-km-radius craters 
at Calypso's six cardinal points. Unlike the case of ejecta from Tethys
(Alvarellos et al. 2005), these ejecta do not get widely scattered,
and instead remain in the vicinity of the Tethys system.
However, no hits on Tethys were observed, 
except for the case of ejecta from Calypso's antapex, 
as expected from their launch speeds. 
In fact most particles come back home to Calypso, 
although a few particles managed to reach the other coorbital moon Telesto; 
all of these are presumably on horseshoe orbits.
Hence, the number of particles reaching Telesto
gives us a \textit{lower} limit on the number of ejecta on horseshoe orbits;
the particles coming back to Calypso may have been on horseshoe or tadpole
orbits, but a simple count cannot differentiate between the two types.

\clearpage
\begin{quotation}
\noindent
Table 4. Fates of ``realistic'' (that is, resulting from a cometary impact) 
ejecta from Calypso after 100,000 years. The ejecta are assumed
to come out of a 1-km-radius crater at the indicated locations.
$N$ is the number of particles with each outcome. 
Percentages have been normalized to the number of particles
which enter Saturn orbit (labeled ``Total''). ``Active'' means
that the particles were still orbiting Saturn at the end of the integration. 
\end{quotation}
\begin{center}
\begin{tabular}{|l|||c|c||c|c||c|c||c|c||c|c|||c|c|}
\hline
&\multicolumn{2}{c||}{N. Pole}&\multicolumn{2}{c||}{S. Pole}&\multicolumn{2}{c||}{Apex}&\multicolumn{2}{c||}{Antisaturn}&\multicolumn{2}{c|||}{Subsaturn}&\multicolumn{2}{c|}{Antapex$^{\dagger}$}\\
\hline
Target  & $N$ & \%  &$N$ & \%   &$N$ & \%  &$N$ & \%  &$N$ & \%  &$N$ & \%  \\
\hline
Tethys  & 0   & 0.0 & 0  &0.0   & 0  &0.0  & 0  & 0.0 &0   & 0.0 & 10 & 5.4 \\
Calypso & 183 &95.3 &185 & 95.9 &174 &92.6 &174 &96.7 &173 &95.1 &149 &81.0 \\
Telesto & 8   & 4.2 & 7  &3.6   & 13 &6.9  & 6  & 3.3 &9   & 4.9 & 24 &13.0 \\
Active  & 1   & 0.5 & 1  &0.5   & 1  &0.5  & 0  & 0.0 &0   & 0.0 & 1  & 0.5 \\
\hline                                                                         
Total   &192 & 100. &193 & 100. &188 & 100.&180 & 100.&182 & 100.&184 &100. \\
\hline
\end{tabular}
\end{center}

{$\dagger$~Antapex initial conditions generated assuming the spallation model
(Melosh 1989) as modified by Zahnle et al. (2008). }

Table 5 likewise lists the fates of ejecta launched from the six cardinal points
on the surface of Telesto. Similar remarks apply as for Table 4.

\begin{quotation}
\noindent
Table 5. Fates of ``realistic'' (that is, resulting from cometary impact) 
ejecta from Telesto after 100,000 years; see Table 4 for explanation.
\end{quotation}
\begin{center}
\begin{tabular}{|l|||c|c||c|c||c|c||c|c||c|c|||c|c|}
\hline
&\multicolumn{2}{c||}{N. Pole}&\multicolumn{2}{c||}{S. Pole}&\multicolumn{2}{c||}{Apex}&\multicolumn{2}{c||}{Antisaturn}&\multicolumn{2}{c|||}{Subsaturn}&\multicolumn{2}{c|}{Antapex$^{\dagger}$}\\
\hline
Target    & $N$ & \% & $N$ & \% & $N$ & \% & $N$ & \% & $N$ & \% & $N$ & \%  \\
\hline
Tethys    & 0   & 0.0& 0   & 0.0& 0   & 0.0& 0   & 0.0& 0   & 0.0& 10  &5.4  \\
Calypso   & 14  &7.2 &12   & 6.3& 19  &10.2&15   &8.4 &11   &6.1 &21   &11.4 \\
Telesto   & 179 &92.3&179  &93.2& 166 &89.2&163  &91.6& 167 &93.3&154  &83.2 \\
Active    & 1   &0.5 & 1   &0.5 & 1   &0.5 & 0   & 0.0& 1   &0.6 & 0   & 0.0 \\
\hline
Total     & 194 &100.&192 &100. & 186 &100.& 178 &100.& 179 &100.& 185 &100. \\
\hline
\end{tabular}
\end{center}

{$\dagger$~Antapex initial conditions generated assuming the spallation model
(Melosh 1989) as modified by Zahnle et al. (2008).}

\clearpage
\section{Discussion and Conclusions}

In this paper we have investigated the fates of ejecta from the Trojan 
companions of Tethys, namely Telesto and Calypso.  In the first
numerical experiment, we systematically investigated the fates of ejecta 
launched from their equatorial regions. 
The initial conditions are translated and rotated from
a topocentric coordinate system (with origin at the Trojan moon's surface 
whence the ejecta are launched) to a saturnicentric system 
and fed to the SWIFT numerical integrator (Levison and Duncan, 1994) 
to propagate the particle trajectories forward in time for 100,000 years. 
As expected, the lowest velocity ejecta cannot fully escape 
the source moon's gravity and fall back onto its surface in less than one day, 
following suborbital trajectories. 

Ejecta do not need to achieve full escape speed to go into orbit about Saturn, 
though; the launch speed $V_0$ only needs to be greater than $V_e^*= \gamma V_e$
(where $V_e$ is the classical escape speed, and $\gamma < 1$; see Eq. 1). 
Particles with $V_0 \approx V_e^*$ or slightly higher 
barely escape the Trojan moon's Hill sphere 
and are confined to tadpole orbits librating about Tethys' $L_4$ point 
(for ejecta from Telesto, leading) or $L_5$ point 
(for ejecta from Calypso, trailing). These particles cannot reach Tethys, 
and instead are eventually re-accreted by Telesto or Calypso respectively. 
An interesting finding is that the removal times of ejecta 
which re-impact Telesto or Calypso within the first dozen years or so 
are discretized to integer multiples of half the tadpole period. 

Somewhat faster ejecta with azimuthal velocities $\gta$ 10 m/s go into horseshoe orbits, 
which enclose not only the $L_4$ and $L_5$ points but also $L_3$; 
these particles still cannot reach Tethys or its Hill sphere 
and are again accreted by either of the Trojan moons. 
Ejecta from Telesto or Calypso reaching tangential velocities 
$\gta$ 40 m/s go into ``passing'' orbits; beyond this point
Tethys accretes most particles, but some are strongly scattered inward 
or outwards towards other saturnian moons in adjacent orbits
such as Enceladus or Dione (on timescales from a few to several hundred years).
It is interesting to note that the timescale for removal of an ejectum
in a horseshoe orbit is usually orders of magnitude longer 
than for ejecta in either tadpole or passing orbits. 

In order to investigate a more realistic distribution of ejecta, 
we assumed an impact producing a 1-km-radius crater at five ``cardinal'' points 
of each Trojan moon (apex, subsaturn, antisaturn, north pole, and south pole). 
The ejection model adopted here is that of Housen et al. (1983), 
consistent with soft regolith: 
the range of ejecta speeds is roughly 5 m/s to 30 m/s. 
As before, initial conditions are integrated for 100,000 years. 
As per our earlier findings, none of these particles can go into passing orbits;
indeed, all of these go into tadpole or horseshoe orbits.
If the former, all particles come back to the source Trojan; 
if the latter, particles can hit either coorbital. However, we noted that 
most (between 89\% and 97\%) of particles come back to the source Trojan moon. 
Most of the rest go into the other Trojan moon, 
with a tiny minority ($<$ 1\%) surviving the 100,000 year integrations. 
Ejecta from the antapex of motion was launched according to a ``spallation'' 
model, with higher speeds of 5 m/s to 56 m/s, so a few percent reached Tethys. 

The ease with which Telesto and Calypso exchange ejecta 
suggests that this recycling may help to maintain 
their similar sizes in the face of impact erosion. 
We intend to test this hypothesis in future work 
on the lopsided Helene-Dione-Polydeuces system. 
A recent preprint (Izidoro {\it et al.} 2010) finds 
that a swarm of planetesimals tadpoling with a proto-satellite 
generally results in the accretion of 1--4 coorbital companions; 
this appears to be consistent with our conclusions as well. 

\section{Acknowledgments}
We would like to acknowledge Patrick Hamill for his contributions, 
and for previewing the manuscript, and to thank R. A. Jacobson
for providing the initial conditions for the Saturn system.
J. L. A. also would like to acknowledge the patience and encouragement
of Alejandra, Joselito, Isabella and Danielito.
This work has made use of NASA's Astrophysics Data System
(ADS located at http://adswww.harvard.edu),
and was supported by NASA's Planetary Geology \& Geophysics Program
through WBS 811073.02.01.03.89.

\newpage

\section{References}

\noindent
Alvarellos, J. L., K. J. Zahnle, A. R. Dobrovolskis, and P. Hamill 2002.
Orbital evolution of impact ejecta from Ganymede. Icarus \textbf{160}, 108-123.
\vspace{11pt}

\noindent
Alvarellos, J. L., K. J. Zahnle, A. R. Dobrovolskis, and P. Hamill 2005.
Fates of satellite ejecta in the Saturn system. Icarus \textbf{178}, 104-123.
\vspace{11pt}

\noindent
Alvarellos, J. L., K. J. Zahnle, A. R. Dobrovolskis, and P. Hamill 2008.
Transfer of mass from Io to Europa and beyond due to cometary impacts.
Icarus \textbf{194}, 636-646.
\vspace{11pt}

\noindent
Arias, E. F., Charlot, P., Feissel, M., and Lestrade, J. F. 1995. 
The extragalactic reference system of the
International Earth Rotation Service, ICRS. 
Astronomy and Astrophysics. \textbf{303}, 604-608.
\vspace{11pt}

\noindent
Bate, R., D. Muller and J. White 1971.
Fundamentals of Astrodynamics.
Dover Publications Inc., New York.
\vspace{11pt}

\noindent
Chapman, C., and W. McKinnon 1986. Cratering of planetary satellites.
In Satellites (J. Burns and M. S. Mathews, Eds.), pp. 492-580.
Univ. of Arizona Press, Tucson.
\vspace{11pt}

\noindent
Danby, J. M. A. 1962.
Fundamentals of Celestial Mechanics.
Willman-Bell Press, Richmond, VA.
\vspace{11pt}

\noindent
Dermott, S. F., and  Murray, C. D. 1981a.
The dynamics of tadpole and horseshoe orbits. I. Theory.
Icarus \textbf{44}, 1--11.
\vspace{11pt}

\noindent
Dermott, S. F., and  Murray, C. D. 1981b.
The dynamics of tadpole and horseshoe orbits.
II. The coorbital satellites of Saturn.
Icarus \textbf{44}, 12--22.
\vspace{11pt}

\noindent
Dobrovolskis, A. R. and J. A. Burns 1980.
Life near the Roche limit: Behavior of ejecta from satellites close to planets.
Icarus \textbf{42}, 422-441.
\vspace{11pt}

\noindent
Dobrovolskis, A. R. and J. J. Lissauer 2004.
The fate of ejecta from Hyperion.
Icarus \textbf{169}, 462-473.
\vspace{11pt}

\noindent
Dobrovolskis, A. R., J. L. Alvarellos and J. J. Lissauer 2007.
Lifetimes of small bodies in planetocentric (or heliocentric) orbits.
Icarus \textbf{188}, 481-505.
\vspace{11pt}

\noindent
Housen, K. R., R. M. Schmidt, and K. A. Holspapple 1983.
Crater ejecta scaling laws: Fundamental forms based on dimensional analysis.
Journal of Geophysical Research \textbf{88}, 2485-2499.
\vspace{11pt}

\noindent
Izidoro, A., O. C. Winter, and M. Tsuchida 2010. 
Coorbital satellites of Saturn: Congenital formation. 
arXiv:10024617v1 (prepared for Monthly Notices of the R.A.S.). 
\vspace{11pt}

\noindent
Jacobson, R. A., Spitale, J., Porco C. C., Beurle, K., 
Cooper, N. J., Evans, M. W., and Murray, C. D. 
Revised orbits of Saturn's small inner satellites. 
The Astronomical Journal \textbf{135}, 261-263.
\vspace{11pt}

\noindent
Kempf, S., U. Beckmann, and J. Schmidt 2010.
How the Enceladus dust plume feeds Saturn's E ring.
Icarus \textbf{206}, 446--457.
\vspace{11pt}

\noindent
Lecacheux, J, Laques, P., Vapillos, L. Auge A. and R. Despiau 1980.
A new satellite of Saturn - Dione B.
Icarus \textbf{43}, 111-115.
\vspace{11pt}

\noindent
Levison, H. F., and M. J. Duncan 1994.
The long-term dynamical behavior of short period comets.
Icarus \textbf{108}, 18-36.
\vspace{11pt}

\noindent
Lissauer, J. J. 1985. Shepherding model for Neptune's arc ring.
Nature \textbf{318}, 544--545.
\vspace{11pt}

\noindent
Melosh, H. J. 1984.
Impact ejection, spallation and the origin of meteorites.
Icarus \textbf{59}, 234-260.
\vspace{11pt}

\noindent
Melosh, H. J. 1985a.
Ejection of rock fragments from planetary bodies.
Geology \textbf{13}, 144-148.
\vspace{11pt}

\noindent
Melosh, H. J. 1985b. Impact cratering mechanics:
Relationship between the shock wave and excavation flow.
Icarus \textbf{62}, 339-343.
\vspace{11pt}

\noindent
Melosh, H. J. 1989. Impact Cratering.
Oxford University Press, New York.
\vspace{11pt}

\noindent
Murray, C.D. and  S. F. Dermott 1999. Solar System Dynamics.
Cambridge University Press, New York.
\vspace{11pt}

\noindent
Pascu D., Harrington R., and P. K. Seidelmann 1980.
Satellites of Saturn. IAU Circ. 3534.
\vspace{11pt}

\noindent
Porco, C. C., and the Cassini Imaging Team 2004.
Satellites and rings of Saturn. IAU Circ. 8432.
\vspace{11pt}

\noindent
Porco, C. C., Thomas, P. C., Weiss, J. W., and Richardson, D. C. 2007. 
Saturn's small inner satellites: Clues to their origins.
Science \textbf{318}, 1602-1607.
\vspace{11pt}

\noindent
Seidelmann P. K., Harrington, R. S., Pascu D., Baum W. A.,
Curie, D. G., Westphal, J. A. and G. E. Danielson 1981.
Saturn  satellite observations and orbits from the 1980 ring plane crossings.
Icarus \textbf{47} 282-287.
\vspace{11pt}

\noindent
Shoemaker, E. M., and R. F. Wolfe 1982.
Cratering time scales for the Galilean satellites.
In Satellites of Jupiter (D. Morrison, Editor), 277--339.
Univ. of Arizona Press, Tucson.
\vspace{11pt}

\noindent
Sinclair, A. T. 1984. 
Perturbations on the orbits of companions of the satellites of Saturn. 
Astronomy \& Astrophysics \textbf{136}, 161--166. 
\vspace{11pt}

\noindent
Smith, B. A., Reitsema, H. J., Fountain, J. W., and S. M. Larson 1980.
Saturn's inner co-orbital satellites.
Bulletin of the American Astronomical Society \textbf{12}, 727.
\vspace{11pt}

\noindent
Smith, B. A. and 26 colleagues 1981.
Encounter with Saturn: Voyager 1 Imaging science results.
Science \textbf{212}, 163-191.
\vspace{11pt}

\noindent
Smith, B. A. and 28 colleagues 1982.
A new look at the Saturn system: The Voyager 2 images.
Science \textbf{215}, 504-537.
\vspace{11pt}

\noindent
Smith, B. A. and 32 colleagues 1986.
Voyager 2 in the Uranian system: Imaging science results.
Science \textbf{233}, 43-64.
\vspace{11pt}

\noindent
Szebehely, V. G. 1967. Theory of Orbits. Academic Press, New York.
\vspace{11pt}

\noindent
Wisdom, J. and M. Holman 1991.
Symplectic maps for the N-body problem.
Astronomical Journal \textbf{102}, 1528-1538.
\vspace{11pt}

\noindent
Zahnle, K., L. Dones, and H. F. Levison 1998.
Cratering rates on the Galilean satellites.
Icarus \textbf{136}, 202-222.
\vspace{11pt}

\noindent
Zahnle, K., P. Schenk, S. Sobieszczyk, L. Dones, and H. F. Levison 2001.
Differential cratering of synchronously rotating satellites by ecliptic comets.
Icarus \textbf{153}, 111-129.
\vspace{11pt}

\noindent
Zahnle, K., P. Schenk, H. Levison and L. Dones 2003.
Cratering rates in the outer solar system.
Icarus \textbf{163}, 263-269.
\vspace{11pt}

\noindent
Zahnle, K. J., J. L. Alvarellos, A. R. Dobrovolskis, and P. Hamill 2008.
Secondary and sesquinary craters on Europa. Icarus \textbf{194}, 660-674.
\vspace{11pt}

\end{document}